\documentclass[12pt]{article}

\usepackage{amsmath}
\usepackage{amssymb}
\usepackage{amsfonts}

\usepackage{color}

\usepackage{graphicx}
\usepackage{citesort}
\graphicspath{{figures/}}
\usepackage{times}

\newcommand{\m}{\mathrm}

\title{
\hspace{-2mm}  Microwave amplification with nanomechanical resonators
}

\author
{ F.\ Massel,$^{1\ast}$ T. T.\ Heikkil\"a,$^{1}$
  J.-M. Pirkkalainen,$^{1}$ S. U. Cho,$^{1}$\\
  H. Saloniemi,$^{2}$ P. Hakonen,$^{1}$ M. A. Sillanp\"a\"a,$^{1}$\\
\normalsize{$^{1}$ Low Temperature Laboratory, Aalto University,}\\
\normalsize{P.O. Box 15100, FI-00076 Aalto, Finland}\\
\normalsize{$^{2}$Microsystems and Nanoelectronics, VTT Technical
  Research Centre of Finland,}\\
\normalsize{ P.O. Box 1000, FI-02044 VTT, Finland}\\
\\
\normalsize{$^\ast$To whom correspondence should be addressed; E-mail:  francesco.massel@aalto.fi.}
}

\begin{document}
\baselineskip24pt
\date{}
\maketitle

\textbf{
  Sensitive measurement of electrical signals is at the heart of
  modern science and technology. According to quantum mechanics, any
  detector or amplifier is required to add a certain amount of noise
  to the signal, equaling at best the energy of quantum
  fluctuations\cite{Caves:1982wq,Haus:2010te}. The quantum limit of
  added noise has nearly been reached with superconducting devices
  which take advantage of nonlinearities in Josephson
  junctions\cite{CastellanosBeltran:2008cg,Bergeal:2010iu}. Here, we
  introduce a new paradigm of amplification of microwave signals with
  the help of a mechanical oscillator. By relying on the radiation
  pressure force on a nanomechanical
  resonator\cite{Braginsky:2001wj,Metzger:2004ei,Kippenberg:2005uo},
  we provide an experimental demonstration and an analytical
  description of how the injection of microwaves induces coherent
  stimulated emission and signal amplification. This scheme, based on
  two linear oscillators, has the advantage of being conceptually and
  practically simpler than the Josephson junction devices, and, at the
  same time, has a high potential to reach quantum limited
  operation. With a measured signal amplification of 25 decibels and
  the addition of 20 quanta of noise, we anticipate near
  quantum-limited mechanical microwave amplification is feasible in
  various applications involving integrated electrical circuits.}

Since the early days of quantum mechanics, the effect of quantum zero
point fluctuations on measurement accuracy has been actively
investigated. When measuring a position $x$ of an object, one
necessarily disturbs its subsequent motion by introducing a
disturbance to the momentum $p$. The imprecision and disturbance are
related by the fundamental limit $\Delta x \Delta p \geq \hbar/2$
owing to the Heisenberg uncertainty principle. A proper compromise
between the two leads to the lowest added noise power per unit
bandwidth $\hbar \omega/2$ which equals the quantum fluctuations of
the system itself at the signal frequency $\omega$. On the other hand,
if only one observable is measured, for example position, or a single
quadrature such as either amplitude or phase of oscillations, noise in
this measurement can be squeezed below the quantum limit at the
expense of increased noise in the other quadrature. In this case, the
amplifier is said to be phase-sensitive.

While most modern transistors operate several orders of magnitude
above the fundamental noise limit, superconducting Josephson junction
parametric
amplifiers\cite{Yurke:1988ih,Andre:1999vh,CastellanosBeltran:2008cg,Bergeal:2010iu},
working near the absolute zero of temperature, have found uses at the
level of only a few added quanta at microwave frequencies.Approaching
the quantum limit with a mechanical amplifier has remained fully
elusive, moreover, there is little work whatsoever on amplifying
electrical signals by mechanical means\cite{Raskin:2000tv}, foremost
due to the typically small electromechanical interaction. In this
work, we describe a way to approach quantum-limited microwave
amplification, now with a mechanical device. Our system consists of a
mechanical resonator affected by radiation pressure forces due to an
electromagnetic field confined in a lithographically patterned
thin-film microwave cavity. Depending on the configuration, it is
capable of either phase-sensitive, or phase-insensitive amplification.

Our system of two coupled linear oscillators forms possibly the
simplest realization of quantum amplification of external signals. The
analysis connects directly to the potential quantum behavior of
macroscopic mechanical
objects\cite{Leggett:2002wc,Marshall:2003kj,OConnell:2010br}, and the
emergence of macroscopic phenomena from the quantum-mechanical laws
governing nature on a microscopic scale. Besides fundamental
significance, the use of mechanical objects as a building block of
low-noise amplification might be advantageous over electrical
realization because of the possibility of obtaining an elementary
physical structure. Ultimately, these devices can be made with
single-crystalline resonant beams or membranes.

\begin{figure}
 \center{
 \includegraphics[width=0.45\textwidth]{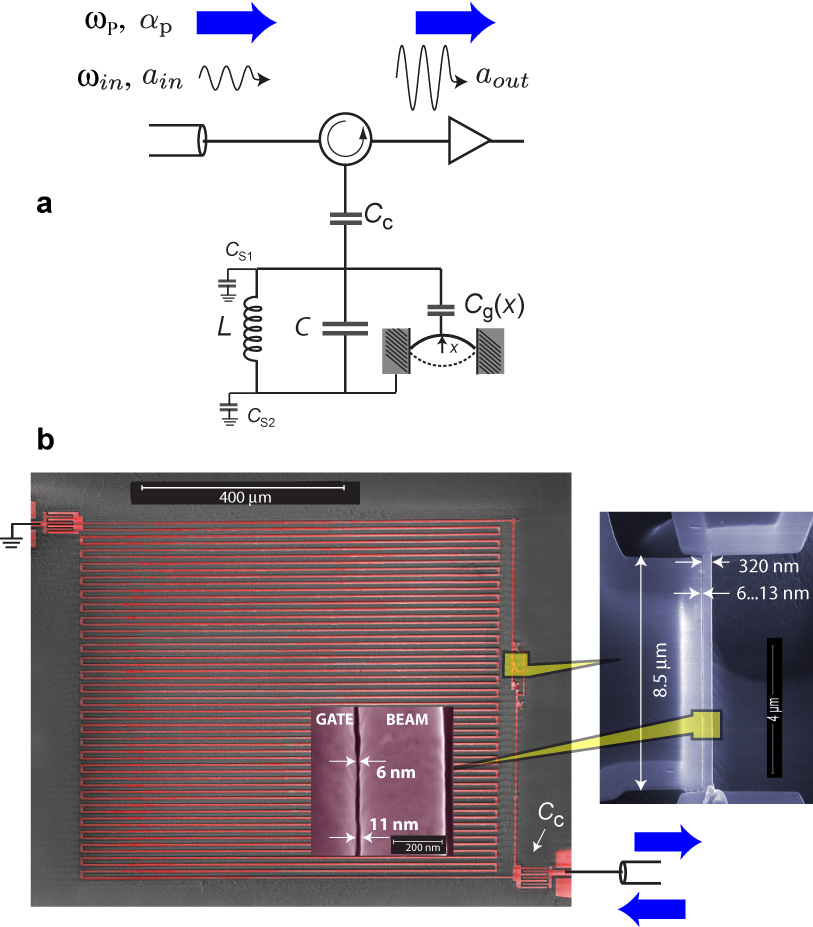}}
  \label{fig:scheme}
  \caption{\textbf{Schematics of the electromechanical microwave
      amplification}. \textbf{a}, The movable capacitance of the
    micromechanical resonator is in parallel to the cavity,
    represented by its lumped element equivalent model. The input
    microwave field is decomposed into a blue-detuned pump coherent
    field $\alpha_p$ at the frequency $\omega_p$, and a signal
    (frequency $\omega_{in}$) + noise $a_{in}=\alpha_{in}+\delta
    a_{in}$. The output signal $a_{out}$ describes the amplified input
    signal.  \textbf{b}, Image of the device is dominated by the
    meandering high-impedance cavity (false color in red), resonating
    at $\omega_c/(2\pi) = 6.982$ GHz. It is connected to coaxial
    cables via a coupling capacitor for operations. The
    micromechanical beam resonator (frequency $\omega_m/(2\pi) = 32.5$
    MHz) couples the ends of the meander via the protrusions from
    either end, and through an 6...15 nm nanometer vacuum
    gap\cite{Sulkko:2010ih} (right, and inset).}
\end{figure}

More specifically, our system involves an on-chip microwave cavity
parametrically interacting with a micromechanical resonator
\cite{Regal:2008di,Sillanpaa:2009hl,Rocheleau:2010jd}, so that the
mechanical motion couples to the frequency of the cavity (see
Fig.~1). In a similar setup, freezing of the mechanical Brownian
motion is expected to take place due to the conversion of mechanical
vibrations into electromagnetic
radiation\cite{Metzger:2004ei,Marquardt:2007dn,WilsonRae:2007jp,Rocheleau:2010jd}. The
interaction gives rise to energy $H_{\m{int}}=-\hbar g \left( n _c
  +\frac{1}{2} \right)x$, where $n_c$ is the number of coherently
driven photons in the cavity, $g = (\omega_c / 2 C) \partial
C_g/ \partial x$ is the electromechanical interaction, and $x$ is the
displacement. The cavity is driven with a strong microwave tone
$\alpha_p$ (hereafter denoted as the pump)
having a frequency $\omega_p \sim
  \omega_c +\omega_m$ which exceeds the resonant frequency $\omega_c$
  of the cavity approximately by the mechanical frequency $\omega_m$,
  see Fig.~2.  The condition $\omega_p > \omega_c$ is referred to as
blue-detuning for the pump with respect to the cavity resonant
frequency, as opposed to the red-detuning condition $\omega_p <
\omega_c$ encountered in the sideband cooling regime both for electro-
and optomechanical systems \cite{Metzger:2004ei,Rocheleau:2010jd}. On
the red-detuned side, the microwave/mechanics coupling generates a net
energy transfer from the mechanical degrees of freedom into the
cavity, hence effectively leading to the cooling of the
mechanics. Contrarily, on the blue-detuning side, the microwave pump
gives rise to an energy transfer into the mechanical degrees of
freedom. This situation can be pictured with the relevant energy
levels as in Fig.~2b: A weak probe signal near the frequency
$\omega_c$ induces stimulated emission of microwave photons,
effectively leading to amplification of the probe signal.

Some details of our actual device are depicted in Fig.~1b. The
micromechanical resonator consists of an aluminum micromechanical beam
resonator having the length $8.5 \, \mu$m, width $320$ nm, which yield
the frequency of the lowest flexural mode $\omega_m/(2\pi) = 32.5$ MHz
and linewidth $\gamma_m \sim 500 ... 1200$ Hz (depending on
temperature). Via an ultranarrow vacuum gap\cite{Sulkko:2010ih}, the
displacement affects the end-to-end capacitance of the cavity, which
is a superconducting microstrip resonator having the natural frequency
$\omega_c \gg \omega_m$ and linewidth $\gamma_c$. In order to obtain
the largest electromechanical coupling, the device was fabricated on a
fused silica substrate. This material has a low dielectric constant
($\epsilon_r \simeq 4$), as compared to, for instance, frequently used
silicon ($\epsilon_r \simeq 12$), which contributes to minimizing the
stray capacitance $C \simeq 18$ fF of the cavity. These measures, and
the remaining equivalent parameters ($L = 21$ nH, $C_{\m{S1}} = 6$ fF,
$C_{\m{S2}} = 12$ fF, $C_{g} = 0.6$ fF, see Fig.~1a), create a strong
electromechanical coupling $g/(2 \pi) \simeq 40$ Hz per phonon, or, in
frequently used units, $g /(2 \pi) = 1.8$ MHz/nm. An interdigital
coupling capacitor $C_c \sim 6$ fF results in the external damping
rate $\gamma_E/(2\pi) = 4.8$ MHz. The total cavity linewidth is
$\gamma_{c} = \gamma_E + \gamma_I \simeq (2 \pi) \times 6.2$ MHz,
where $\gamma_I/(2\pi) = 1.4$ MHz is due to internal losses.

\begin{figure}[!ht]
 \label{fig:mechanism}
 \center{
 \includegraphics[width=0.45\textwidth]{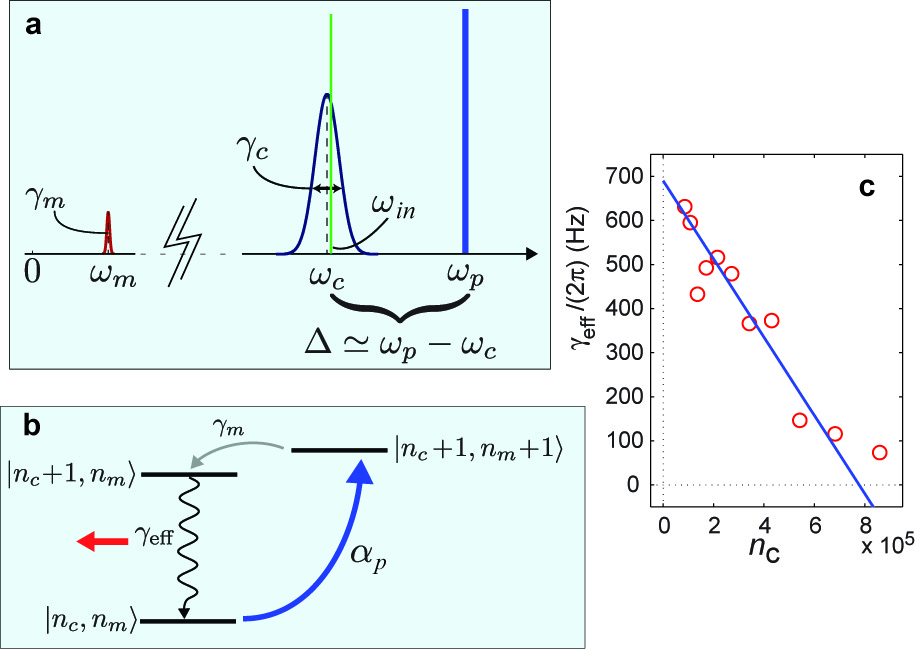}
 }
 \caption{\textbf{Amplification mechanism}. \textbf{a}, Explanation of
   the various frequencies involved: The cavity is driven by a (pump,
   $\alpha_p$) coherent field oscillating at
   $\omega_p=\omega_c+\Delta$, where $\omega_c$ is the cavity resonant
   frequency, and $\Delta \simeq \omega_m\ll \omega_c$. The weak input
   signal $a_{\m{in}}$ to be amplified is applied near the cavity
   resonance, such that $\omega_{\m{in}} = \omega_p -
   \omega_m$. \textbf{b}, The physics of the electromechanical
   amplifier can be intuitively understood by considering a block of
   three energy levels in the system. The (blue-detuned) pump
   $\alpha_{p}$ induces a transition between a state characterized by
   $n_c$ cavity photons and $n_m$ mechanical quanta and a state with
   $n_c+1$ and $n_m+1$. A key role in the amplification process is
   played by the effective damping $\gamma_{\m{eff}}$, which in the
   simplified scheme presented here represents the effective lifetime
   for the cavity photons, thus modelling the parametric-down
   conversion of the pump photons to the cavity resonant
   frequency. \textbf{c}, The damping decreases towards a higher pump
   field, while below $\gamma_{\m{eff}} = 0$, instability and
   parametric oscillations take place. The circles are fitted from the
   measured mechanical peak in the output spectrum, and the solid line
   is theory.}
\end{figure}

The theoretical framework suitable for description of the
amplification is closely related to the methodology used to describe
the sideband cooling in optomechanical systems
\cite{Mancini:1994vf,Marquardt:2007dn,WilsonRae:2007jp,Genes:2008hr,Walls:1105914,Clerk:2010dh}. We
describe the system in terms of quantum Langevin equations with the
aim of analyzing the effect of the pumping on the signal, and
especially to detail the effects of different types of fluctuations
coupling to the system. The latter are caused by the quantum and
thermal fluctuations related to the input signal, the cavity, and the
mechanical resonator. In general, the parametric coupling between the
cavity and the mechanical resonator gives rise to the possibility of
squeezing \cite{Rugar:1991he,Woolley:2008ds}, and hence to back-action
evading measurements.

\begin{figure}
 \label{fig:gain}
 \center{
 \includegraphics[width=0.9\textwidth]{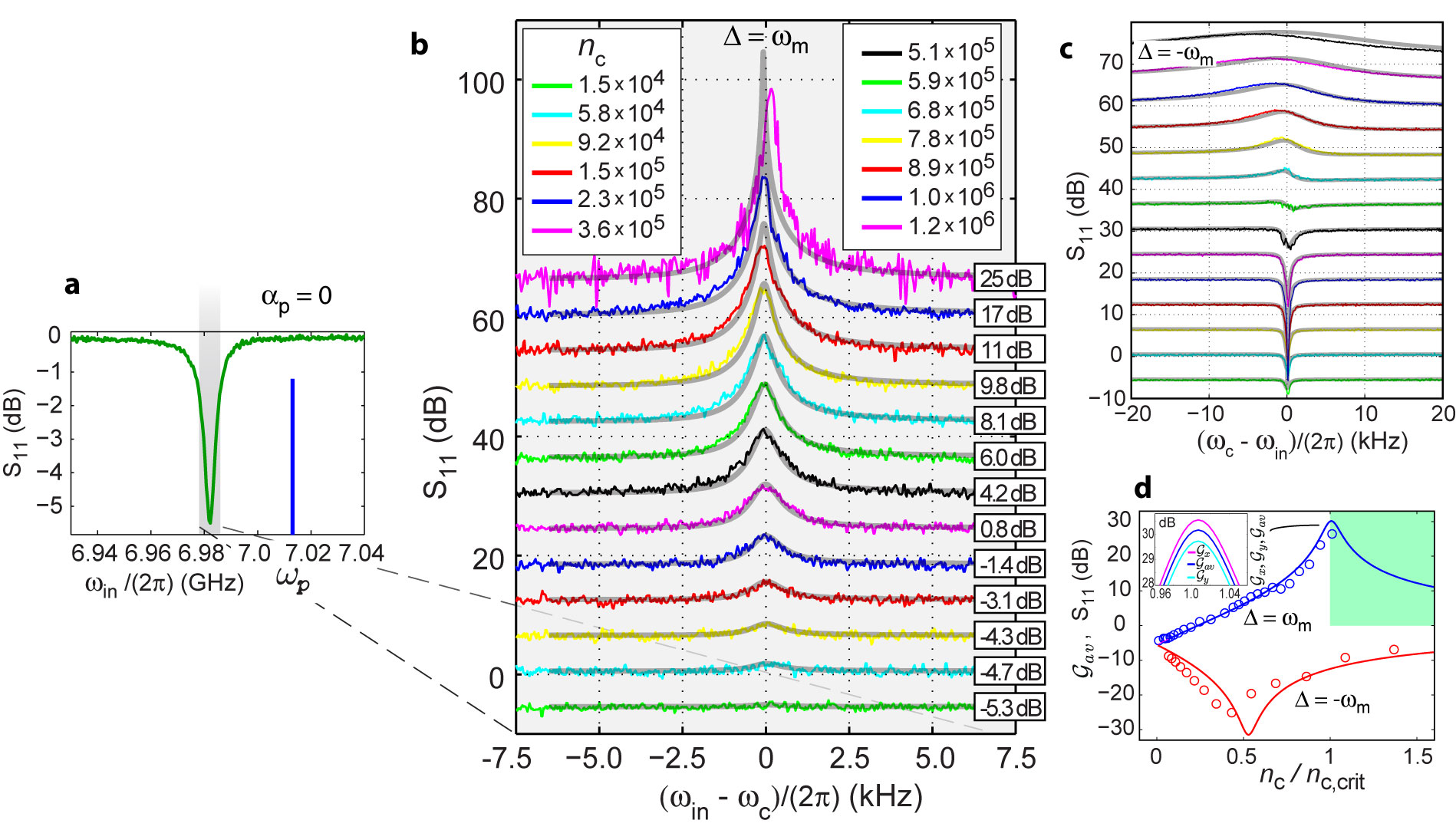}
 }
 \caption{\textbf{Amplifier gain}. \textbf{a}, Measured reflection
   magnitude of the signal microwave over a large span about the
   cavity resonance without applying the pump. \textbf{b}, Measured
   gain (colored lines) in a narrow window (schematically indicated by
   gray in A) for the pump occupation $n_c$ increasing by 2 dB per
   curve from bottom to top. The curves are displaced vertically for
   clarity by 3 dB. The baseline of each curve roughly corresponds to
   the cavity absorption of -5.5 dB. The maximum gain is quoted for
   each curve. The gray curves are theoretical predictions for the
   gain $\mathcal{G}_{av}$ averaged over quadratures. \textbf{c}, As
   panel B, but with \emph{red} detuned pumping. Here,
   de-amplification of the signal field (sharp dip) is observed. $n_c$
   increases by 2 dB per curve from bottom ($4.6 \times 10^4$) to top.
   \textbf{d}, Calculated average gain (at $\omega_p-\omega_m$), and
   the quadrature gains $\mathcal{G}_x$, $\mathcal{G}_y$ (inset) as a
   function of the scaled cavity occupancy. The blue traces correspond
   to amplification ($\Delta > 0$), and red to de-amplification
   ($\Delta < 0$). The shaded region is the unstable regime (see
   text). The experimental points are extracted from B and C.}
\end{figure}

The detailed theoretical analysis (see supplementary information)
gives the explicit value of the gain in each preferred quadrature
$\mathcal{G}_x$, $\mathcal{G}_y$ and the average gain
$\mathcal{G}_{av}=\frac{1}{2}\left(\mathcal{G}_x+\mathcal{G}_y\right)$. The
expression for the gains are well approximated by
\begin{align}
  \label{eq:gain}
  \mathcal{G}_{x,y}=4\left|\Gamma_M (\omega) \right|^2
  \left(\frac{\gamma_E}{\gamma_c}\right)^2
  \left(\sqrt{1+\left(\frac{\gamma_c}{4\Delta}\right)^2}\pm
    \frac{\gamma_c}{\Delta}\right)^2 \, ,
\end{align}
where the upper (lower) sign corresponds to $\mathcal{G}_x$
($\mathcal{G}_y$).

The key role in the microwave amplification is played by the effective
mechanical damping
$\gamma_{\m{eff}}=\gamma_m-\delta\gamma_{\m{eff}}(\omega,n_c)$ (see
Fig.~2c and supplementary information). The value of
$\delta\gamma_{\m{eff}}$ can be tuned by the pump. In particular,
blue-detuned drive corresponds to a sizable reduction of the
mechanical damping which is directly related to the signal
amplification mechanism described by the factor
$\Gamma_M(\omega)=\left(\omega_m^2-\omega^2-i \gamma_m \omega\right)/
\left(\omega_{\m{eff}}^2-\omega^2-i \gamma_{\m{eff}}\omega\right)$,
where $\omega_{\m{eff}} \simeq \omega_m$. Equation~\eqref{eq:gain} and
the definition of $\gamma_{\m{eff}}$ allow to establish the optimal
value for the pumped occupancy as $n_{c,\m{crit}}=\gamma_c\gamma_m/(4
g^2 x_0^2)$ associated with the maximum gain
$\mathcal{G}_{av}(\omega=\omega_m)\simeq 4\left(
  4\Delta/\gamma_c\right)^2$. Above this threshold, $\gamma_{\m{eff}}
\rightarrow 0$, and the coupled system becomes unstable
\cite{Braginsky:2001wj,Carr:2000uv,Kippenberg:2005uo,Arcizet:2006hv}.

Indicated by the unequal $\mathcal{G}_x$ and $\mathcal{G}_y$, the
amplifier will portray a variable amount of squeezing. In particular,
in the so-called good cavity limit $\frac{\gamma_c}{\Delta} \ll 1$
used in the present experiment, the squeezing is expected to be
insignificant, and the amplifier is hence behaving as a typical
phase-insensitive amplifier characterized by the average gain
$\mathcal{G}_{av}$. However, it is noteworthy that by varying the
parameters towards the bad cavity limit $\frac{\gamma_c}{\Delta} \gg
1$, one may achieve strong squeezing and, consequently, noise-free
amplification in one quadrature.

We used a dilution refrigerator with temperature stabilized between
350...25 mK in order to carry out the measurements. The pump and
signal tones are combined at room temperature using a power
splitter. Inside the cryostat, the incoming irradiation is attenuated
by $43 \pm 1.5$ dB. The uncertainty in the cryogenic attenuation sets
the relatively large error bars for $n_c$. The signals reflected from
the device chip are directed towards conventional cryogenic
amplifier. Following amplification, we used a sharp low-pass filter to
remove the strong pump microwave, since we only consider the signal
frequency in the processing. The signal is recorded with a vector
network analyzer, or, for noise measurements, fed into a spectrum
analyzer. Although it is possible to distinguish the different
quadrature gains $\mathcal{G}_{x}$, $\mathcal{G}_{y}$, for the moment
we only consider the phase-insensitive gain $\mathcal{G}_{av}$. More
details of the practical setup are given in the supplementary
information.

At the lowest pump power $n_c \ll n_{c,\m{crit}}$, we observe an
attenuation of the reflected signal according to the cavity absorption
at the applied \emph{signal} frequency $\omega_{in} \sim 6.82$ GHz,
for instance, $\sim -6$ dB around $\omega_{in} \sim \omega_c$
(Fig.~3a). Towards higher $n_c$, a mechanical peak at $\omega_p
-\omega_{in} \simeq \omega_m$ indicates a decreased absorption. A
vanishing absorption is observed when $ 2 \gamma_m /\gamma_{\m{eff}}
\sim \gamma_E /\gamma_c$, i.e., when reduced damping balances external
losses. This marks an onset of a type of electromechanically induced
microwave transparency. Earlier work on the effect
\cite{Weis:2010ci,Teufel:2011ha} used the opposite red-detuned pump
conditions. Blue pumping, on the other hand, has the capability of
transferring energy to the cavity frequency, and therefore, we observe
also amplification of the signal tone, Fig.~3b.

The theoretical prediction, overlaid on the experimental data in
Fig.~3b, shows a remarkable agreement with the measurement, with the
only adjustable parameter being $n_c$, whose values, nevertheless, are
independently known within a factor of 2. The maximum gain of 26 dB is
observed close to the instability, however, with markedly increased
amount of fluctuations, typical of parametric systems. The useful
range of gain where noise is not increased, is up to about 15 dB in
this measurement. The operation frequency of the amplifier may be
tuned roughly by the amount of the cavity linewidth by detuning the
pump slightly off from the exact blue sideband condition. While the
data was measured with a small input signal corresponding to about 0.3
signal photons in cavity, a favorable property of the mechanical
amplification is its high dynamic range, namely, amplification is
observed at least up to $10^4$ signal photons. This high power
handling capability is in striking contrast to the Josephson devices
which work in the single-quantum regime.

 We can also confirm the theory predictions by inverting the pump
 frequency to the red sideband, i.e., $\Delta = -\omega_m$, see
 Fig.~3c. In analog to the enhanced damping under these
 conditions leading to sideband cooling, we observe de-amplification
 of the input signal. At even stronger pump $n_c > n_{c, \m{crit}}$,
 decreased absorption indicates that the eigenmode splitting
\cite{Teufel:2011ha} starts to set in. The observed
 gains (Fig.~3d) as a function of the pumped occupancy are
 in a good agreement with theory.

 Finally, an amplifier should be characterized by its added noise
 which is commonly referred to the input signal. Because of its simple
 structure, we expect the mechanical amplifier to be less influenced
 by the strong $1/f$ electrical flicker noise typical of nonlinear
 transistor or Josephson junction devices. What remain as noise
 sources are thermal fluctuations, and, ultimately, quantum
 fluctuations in the number of quanta of both cavity and the
 mechanics. In the case of the optimal gain, the quadrature-averaged
 added noise at the cavity frequency is
\begin{align}
  \label{eq:addnoise}
    \begin{split}
  n_{\rm add}= & \frac{\gamma_I}{\gamma_{c}}\left(n^T_{c}+\frac{1}{2}\right)
  + \frac{\gamma_c}{\gamma_E}\left(n_m+\frac{1}{2}\right) \\
  & \geq   \frac{1}{2}\left[1-\left(\mathcal{G}_x\mathcal{G}_y\right)^{-1/2}\right]\simeq \frac{1}{2}.
    \end{split}
 \end{align}
 Here, $n^T_{c}$ and $n_m$ represent the finite number of cavity
 photons associated with the internal losses, and the number of
 mechanical quanta due to the thermal fluctuations in the mechanical
 bath, respectively. The quantum limit $n_{\rm add}= 1/2$ with large
 gain may thus be reached if information is not lost inside the
 system, that is, if external dissipation of the cavity related to the
 measurement dominates over internal losses. While in cryogenic
 experiments typically $n^T_c \simeq 0$ due to gigahertz-range cavity
 frequencies, standard mechanical frequencies $\omega_m/(2\pi)
 \lesssim 50$ MHz tend to pose a practical limit for noise $n_{\rm
   add} > n_m \gtrsim 10$.

 We measured the added noise by comparing it to a known noise
 source. Here, the noise floor is set by the effective noise
 temperature of the system, approximately 6...7 K. We worked at a
 temperature of 30 mK, and used a weak input signal as a marker (see
 supplementary information). We obtained a slight 0.6 dB improvement
 to the signal-to-noise ratio, which corresponds to 20 added noise
 quanta at the signal frequency of 7 GHz. This finding agrees
 remarkably well with the ideal prediction equaling the thermal phonon
 number, and shows that no extra noise appeared in the process. Hence,
 an even further improved performance closer to the quantum limit
 looks promising.

 We have shown that interaction of a micromechanical device and
 radiation pressure can be used for amplifying weak electrical
 signals. This finding opens up new perspectives for an alternative to
 the conventional electrical microwave amplifiers, and may facilitate
 radiation detection in the difficult terahertz band. From a
 theoretical point of view, the setup represents one of the simplest
 realizations of a quantum amplifier leading to operation at the noise
 limit set by the Heisenberg uncertainty principle. In the first
 proof-of-principle device, the mechanical amplifier showed no extra
 added noise beyond that predicted by the ideal theoretical model. For
 a practical application, the frequency band can be made variable over
 a larger span by using tunable
 cavities\cite{PalaciosLaloy:2008ef,CastellanosBeltran:2008cg}. An
 even higher electromechanical coupling, for instance, by the use of
 piezoelectric materials\cite{OConnell:2010br}, can allow for an
 increase of the band via an engineered increase of damping. At a
 higher mechanical frequency in the GHz range, or with the help of
 pre-cooling by opposite pumping at even high temperatures, we can
 foresee near-quantum limited operation.

\textbf{Acknowledgements} We would like to thank Sorin Paraoanu for
useful discussions. This work was supported by the Academy of Finland,
and by the European Research Council (grants No. 240362-Heattronics and
240387-NEMSQED) and EU-FP7-NMP-246026.

\textbf{Author Contributions} F.M. and T.H. developed the
theory. J.-M.P. and S.U.C. contributed to design and fabrication of
the samples, and cryogenic setup. H.S. made the samples. P.H. and
M.S. designed the experimental setup. M.S. initiated the work and
carried out the measurements.

\textbf{Author Information} The authors declare no competing financial
interests. Correspondence and requests for materials should be
addressed to F.M. (francesco.massel@aalto.fi).

\bibliographystyle{naturemag}

\begin{thebibliography}{10}
\expandafter\ifx\csname url\endcsname\relax
  \def\url#1{\texttt{#1}}\fi
\expandafter\ifx\csname urlprefix\endcsname\relax\def\urlprefix{URL }\fi
\providecommand{\bibinfo}[2]{#2}
\providecommand{\eprint}[2][]{\url{#2}}

\bibitem{Caves:1982wq}
\bibinfo{author}{Caves, C.~M.}
\newblock \bibinfo{title}{{Quantum limits on noise in linear-amplifiers}}.
\newblock \emph{\bibinfo{journal}{Phys. Rev. D}} \textbf{\bibinfo{volume}{26}},
  \bibinfo{pages}{1817--1839} (\bibinfo{year}{1982}).

\bibitem{Haus:2010te}
\bibinfo{author}{Haus, H.~A.}
\newblock \emph{\bibinfo{title}{{Electromagnetic Noise and Quantum Optical
  Measurements (Advanced Texts in Physics)}}} (\bibinfo{publisher}{Springer},
  \bibinfo{year}{2000}).

\bibitem{CastellanosBeltran:2008cg}
\bibinfo{author}{Castellanos-Beltran, M.~A.}, \bibinfo{author}{Irwin, K.~D.},
  \bibinfo{author}{Hilton, G.~C.}, \bibinfo{author}{Vale, L.~R.} \&
  \bibinfo{author}{Lehnert, K.~W.}
\newblock \bibinfo{title}{{Amplification and squeezing of quantum noise with a
  tunable Josephson metamaterial}}.
\newblock \emph{\bibinfo{journal}{Nat. Phys.}} \textbf{\bibinfo{volume}{4}},
  \bibinfo{pages}{929--931} (\bibinfo{year}{2008}).

\bibitem{Bergeal:2010iu}
\bibinfo{author}{Bergeal, N.} \emph{et~al.}
\newblock \bibinfo{title}{{Phase-preserving amplification near the quantum
  limit with a Josephson ring modulator}}.
\newblock \emph{\bibinfo{journal}{Nature}} \textbf{\bibinfo{volume}{465}},
  \bibinfo{pages}{64--68} (\bibinfo{year}{2010}).

\bibitem{Braginsky:2001wj}
\bibinfo{author}{Braginsky, V.~B.}, \bibinfo{author}{Strigin, S.~E.} \&
  \bibinfo{author}{Vyatchanin, S.~P.}
\newblock \bibinfo{title}{{Parametric oscillatory instability in Fabry-Perot
  interferometer}}.
\newblock \emph{\bibinfo{journal}{Phys. Lett. A}}
  \textbf{\bibinfo{volume}{287}}, \bibinfo{pages}{331--338}
  (\bibinfo{year}{2001}).

\bibitem{Metzger:2004ei}
\bibinfo{author}{Metzger, C.~H.} \& \bibinfo{author}{Karrai, K.}
\newblock \bibinfo{title}{{Cavity cooling of a microlever}}.
\newblock \emph{\bibinfo{journal}{Nature}} \textbf{\bibinfo{volume}{432}},
  \bibinfo{pages}{1002--1005} (\bibinfo{year}{2004}).

\bibitem{Kippenberg:2005uo}
\bibinfo{author}{Kippenberg, T.~J.}, \bibinfo{author}{Rokhsari, H.},
  \bibinfo{author}{Carmon, T.}, \bibinfo{author}{Scherer, A.} \&
  \bibinfo{author}{Vahala, K.~J.}
\newblock \bibinfo{title}{{Analysis of radiation-pressure induced mechanical
  oscillation of an optical microcavity}}.
\newblock \emph{\bibinfo{journal}{Phys. Rev. Lett.}}
  \textbf{\bibinfo{volume}{95}}, \bibinfo{pages}{--} (\bibinfo{year}{2005}).

\bibitem{Yurke:1988ih}
\bibinfo{author}{Yurke, B.} \emph{et~al.}
\newblock \bibinfo{title}{{Observation of 4.2-K equilibrium-noise squeezing via
  a Josephson-parametric amplifier}}.
\newblock \emph{\bibinfo{journal}{Phys. Rev. Lett.}}
  \textbf{\bibinfo{volume}{60}}, \bibinfo{pages}{764--767}
  (\bibinfo{year}{1988}).

\bibitem{Andre:1999vh}
\bibinfo{author}{Andr{\'e}, M.~O.}, \bibinfo{author}{M{\"u}ck, M.},
  \bibinfo{author}{Clarke, J.}, \bibinfo{author}{Gail, J.} \&
  \bibinfo{author}{Heiden, C.}
\newblock \bibinfo{title}{{Radio-frequency amplifier with tenth-kelvin noise
  temperature based on a microstrip direct current superconducting quantum
  interference device}}.
\newblock \emph{\bibinfo{journal}{Appl. Phys. Lett.}}
  \textbf{\bibinfo{volume}{75}}, \bibinfo{pages}{698--700}
  (\bibinfo{year}{1999}).

\bibitem{Raskin:2000tv}
\bibinfo{author}{Raskin, J.~P.}, \bibinfo{author}{Brown, A.~R.},
  \bibinfo{author}{Khuri-Yakub, B.~T.} \& \bibinfo{author}{Rebeiz, G.~M.}
\newblock \bibinfo{title}{{A novel parametric-effect MEMS amplifier}}.
\newblock \emph{\bibinfo{journal}{J. Microelectromech. S.}}
  \textbf{\bibinfo{volume}{9}}, \bibinfo{pages}{528--537}
  (\bibinfo{year}{2000}).

\bibitem{Leggett:2002wc}
\bibinfo{author}{Leggett, A.~J.}
\newblock \bibinfo{title}{{Testing the limits of quantum mechanics: motivation,
  state of play, prospects}}.
\newblock \emph{\bibinfo{journal}{J. Phys-Condens. Mat.}}
  \textbf{\bibinfo{volume}{14}}, \bibinfo{pages}{R415--R451}
  (\bibinfo{year}{2002}).

\bibitem{Marshall:2003kj}
\bibinfo{author}{Marshall, W.}, \bibinfo{author}{Simon, C.},
  \bibinfo{author}{Penrose, R.} \& \bibinfo{author}{Bouwmeester, D.}
\newblock \bibinfo{title}{{Towards Quantum Superpositions of a Mirror}}.
\newblock \emph{\bibinfo{journal}{Phys. Rev. Lett.}}
  \textbf{\bibinfo{volume}{91}}, \bibinfo{pages}{130401}
  (\bibinfo{year}{2003}).

\bibitem{OConnell:2010br}
\bibinfo{author}{O'Connell, A.~D.} \emph{et~al.}
\newblock \bibinfo{title}{{Quantum ground state and single-phonon control of a
  mechanical resonator}}.
\newblock \emph{\bibinfo{journal}{Nature}} \textbf{\bibinfo{volume}{464}},
  \bibinfo{pages}{697--703} (\bibinfo{year}{2010}).

\bibitem{Regal:2008di}
\bibinfo{author}{Regal, C.~A.}, \bibinfo{author}{Teufel, J.~D.} \&
  \bibinfo{author}{Lehnert, K.~W.}
\newblock \bibinfo{title}{{Measuring nanomechanical motion with a microwave
  cavity interferometer}}.
\newblock \emph{\bibinfo{journal}{Nat. Phys.}} \textbf{\bibinfo{volume}{4}},
  \bibinfo{pages}{555--560} (\bibinfo{year}{2008}).

\bibitem{Sillanpaa:2009hl}
\bibinfo{author}{Sillanp{\"a}{\"a}, M.~A.}, \bibinfo{author}{Sarkar, J.},
  \bibinfo{author}{Sulkko, J.}, \bibinfo{author}{Muhonen, J.} \&
  \bibinfo{author}{Hakonen, P.~J.}
\newblock \bibinfo{title}{{Accessing nanomechanical resonators via a fast
  microwave circuit}}.
\newblock \emph{\bibinfo{journal}{Appl. Phys. Lett.}}
  \textbf{\bibinfo{volume}{95}}, \bibinfo{pages}{011909}
  (\bibinfo{year}{2009}).

\bibitem{Rocheleau:2010jd}
\bibinfo{author}{Rocheleau, T.} \emph{et~al.}
\newblock \bibinfo{title}{{Preparation and detection of a mechanical resonator
  near the ground state of motion}}.
\newblock \emph{\bibinfo{journal}{Nature}} \textbf{\bibinfo{volume}{463}},
  \bibinfo{pages}{72--75} (\bibinfo{year}{2010}).

\bibitem{Marquardt:2007dn}
\bibinfo{author}{Marquardt, F.}, \bibinfo{author}{Chen, J.~P.},
  \bibinfo{author}{Clerk, A.~A.} \& \bibinfo{author}{Girvin, S.~M.}
\newblock \bibinfo{title}{{Quantum theory of cavity-assisted sideband cooling
  of mechanical motion}}.
\newblock \emph{\bibinfo{journal}{Phys. Rev. Lett.}}
  \textbf{\bibinfo{volume}{99}}, \bibinfo{pages}{093902}
  (\bibinfo{year}{2007}).

\bibitem{WilsonRae:2007jp}
\bibinfo{author}{Wilson-Rae, I.}, \bibinfo{author}{Nooshi, N.},
  \bibinfo{author}{Zwerger, W.} \& \bibinfo{author}{Kippenberg, T.~J.}
\newblock \bibinfo{title}{{Theory of ground state cooling of a mechanical
  oscillator using dynamical backaction}}.
\newblock \emph{\bibinfo{journal}{Phys. Rev. Lett.}}
  \textbf{\bibinfo{volume}{99}}, \bibinfo{pages}{093901}
  (\bibinfo{year}{2007}).

\bibitem{Sulkko:2010ih}
\bibinfo{author}{Sulkko, J.} \emph{et~al.}
\newblock \bibinfo{title}{{Strong Gate Coupling of High-Q Nanomechanical
  Resonators}}.
\newblock \emph{\bibinfo{journal}{Nano Lett.}} \textbf{\bibinfo{volume}{10}},
  \bibinfo{pages}{4884--4889} (\bibinfo{year}{2010}).

\bibitem{Mancini:1994vf}
\bibinfo{author}{Mancini, S.} \& \bibinfo{author}{Tombesi, P.}
\newblock \bibinfo{title}{{Quantum-noise reduction by radiation pressure}}.
\newblock \emph{\bibinfo{journal}{Phys. Rev. A}} \textbf{\bibinfo{volume}{49}},
  \bibinfo{pages}{4055--4065} (\bibinfo{year}{1994}).

\bibitem{Genes:2008hr}
\bibinfo{author}{Genes, C.}, \bibinfo{author}{Vitali, D.},
  \bibinfo{author}{Tombesi, P.}, \bibinfo{author}{Gigan, S.} \&
  \bibinfo{author}{Aspelmeyer, M.}
\newblock \bibinfo{title}{{Ground-state cooling of a micromechanical
  oscillator: Comparing cold damping and cavity-assisted cooling schemes}}.
\newblock \emph{\bibinfo{journal}{Phys. Rev. A}} \textbf{\bibinfo{volume}{77}},
  \bibinfo{pages}{033804} (\bibinfo{year}{2008}).

\bibitem{Walls:1105914}
\bibinfo{author}{Walls, D.~F.} \& \bibinfo{author}{Milburn, G.~J.}
\newblock \emph{\bibinfo{title}{{Quantum Optics}}}
  (\bibinfo{publisher}{Springer}, \bibinfo{address}{Berlin},
  \bibinfo{year}{2007}).

\bibitem{Clerk:2010dh}
\bibinfo{author}{Clerk, A.~A.}, \bibinfo{author}{Devoret, M.~H.},
  \bibinfo{author}{Girvin, S.~M.}, \bibinfo{author}{Marquardt, F.} \&
  \bibinfo{author}{Schoelkopf, R.~J.}
\newblock \bibinfo{title}{{Introduction to quantum noise, measurement, and
  amplification}}.
\newblock \emph{\bibinfo{journal}{Rev. Mod. Phys.}}
  \textbf{\bibinfo{volume}{82}}, \bibinfo{pages}{1155--1208}
  (\bibinfo{year}{2010}).

\bibitem{Rugar:1991he}
\bibinfo{author}{Rugar, D.} \& \bibinfo{author}{Gr{\"u}tter, P.}
\newblock \bibinfo{title}{{Mechanical parametric amplification and
  thermomechanical noise squeezing}}.
\newblock \emph{\bibinfo{journal}{Phys. Rev. Lett.}}
  \textbf{\bibinfo{volume}{67}}, \bibinfo{pages}{699--702}
  (\bibinfo{year}{1991}).

\bibitem{Woolley:2008ds}
\bibinfo{author}{Woolley, M.~J.}, \bibinfo{author}{Doherty, A.~C.},
  \bibinfo{author}{Milburn, G.~J.} \& \bibinfo{author}{Schwab, K.~C.}
\newblock \bibinfo{title}{{Nanomechanical squeezing with detection via a
  microwave cavity}}.
\newblock \emph{\bibinfo{journal}{Phys. Rev. A}} \textbf{\bibinfo{volume}{78}},
  \bibinfo{pages}{062303} (\bibinfo{year}{2008}).

\bibitem{Carr:2000uv}
\bibinfo{author}{Carr, D.~W.}, \bibinfo{author}{Evoy, S.},
  \bibinfo{author}{Sekaric, L.}, \bibinfo{author}{Craighead, H.~G.} \&
  \bibinfo{author}{Parpia, J.~M.}
\newblock \bibinfo{title}{{Parametric amplification in a torsional
  microresonator}}.
\newblock \emph{\bibinfo{journal}{Appl. Phys. Lett.}}
  \textbf{\bibinfo{volume}{77}}, \bibinfo{pages}{1545--1547}
  (\bibinfo{year}{2000}).

\bibitem{Arcizet:2006hv}
\bibinfo{author}{Arcizet, O.}, \bibinfo{author}{Cohadon, P.~F.},
  \bibinfo{author}{Briant, T.}, \bibinfo{author}{Pinard, M.} \&
  \bibinfo{author}{Heidmann, A.}
\newblock \bibinfo{title}{{Radiation-pressure cooling and optomechanical
  instability of a micromirror}}.
\newblock \emph{\bibinfo{journal}{Nature}} \textbf{\bibinfo{volume}{444}},
  \bibinfo{pages}{71--74} (\bibinfo{year}{2006}).

\bibitem{Weis:2010ci}
\bibinfo{author}{Weis, S.} \emph{et~al.}
\newblock \bibinfo{title}{{Optomechanically Induced Transparency}}.
\newblock \emph{\bibinfo{journal}{Science}} \textbf{\bibinfo{volume}{330}},
  \bibinfo{pages}{1520--1523} (\bibinfo{year}{2010}).

\bibitem{Teufel:2011ha}
\bibinfo{author}{Teufel, J.~D.} \emph{et~al.}
\newblock \bibinfo{title}{{Circuit cavity electromechanics in the
  strong-coupling regime}}.
\newblock \emph{\bibinfo{journal}{Nature}} \textbf{\bibinfo{volume}{471}},
  \bibinfo{pages}{204--208} (\bibinfo{year}{2011}).

\bibitem{PalaciosLaloy:2008ef}
\bibinfo{author}{Palacios-Laloy, A.} \emph{et~al.}
\newblock \bibinfo{title}{{Tunable resonators for quantum circuits}}.
\newblock \emph{\bibinfo{journal}{J. Low Temp. Phys.}}
  \textbf{\bibinfo{volume}{151}}, \bibinfo{pages}{1034--1042}
  (\bibinfo{year}{2008}).

\end{thebibliography}


\end{document}



\baselineskip24pt

\date{}
\maketitle

\section{Description of the experiment}
\label{sec:exp}

\subsection{Sample fabrication}

In order to minimize the stray capacitance $C$ of the cavity, the
device was fabricated on a fused silica (SiO$_2$, glass) substrate,
which has a low dielectric constant ($\epsilon_r \simeq 4$), as
compared to, for instance, silicon ($\epsilon_r \simeq 12$).

Both the cavity and the structures for the beam were made in a single
e-beam lithography step, followed by evaporation of 150 nm
aluminum. In order to suspend the beams, the substrate was etched by
the use of HF vapor etcher, for 500 seconds at a pressure of 150
torr. The use of HF vapor instead of liquid oxide etchant is necessary
in order to avoid damaging the aluminum film. The depth of the roughly
isotropic etch was about 700 nm.

The mechanical beams were defined by Focused Ion Beam (FIB) etching,
as in Ref.~\cite{Sulkko:2010ih}. In order to create uniform 9-12 nm vacuum slit
over the whole length $L = 8.5 \, \mu$m of the beam, we used low
gallium ion currents of 1.5 pA and 75\% exposure overlap in a single
cutting pass. In order to avoid charge accumulation due to the
insulating substrate, all the structures were kept galvanically
short-circuited and connected to ground until the very end of
fabrication.

\subsection{Cavity design and characterization}
\label{sec:cavity}

The cavity was designed and fabricated such that it would have a high
critical current in order to enable a high drive $n_c \gtrsim 10^8$,
and as small stray capacitance $C$ as possible. The first requirement
suggests to fabricate it in a single lithography step. The low
dielectric constant ($\epsilon_r \simeq 4$) of the substrate
contributes to a low stray capacitance, moreover, the roughly
isotropic release etch for the beam, which partially suspended also
the cavity, finally contributed such that $\epsilon_r \sim 3$ for the
final structure.

\begin{figure}[ht]
\centering
\includegraphics[width=0.9\textwidth]{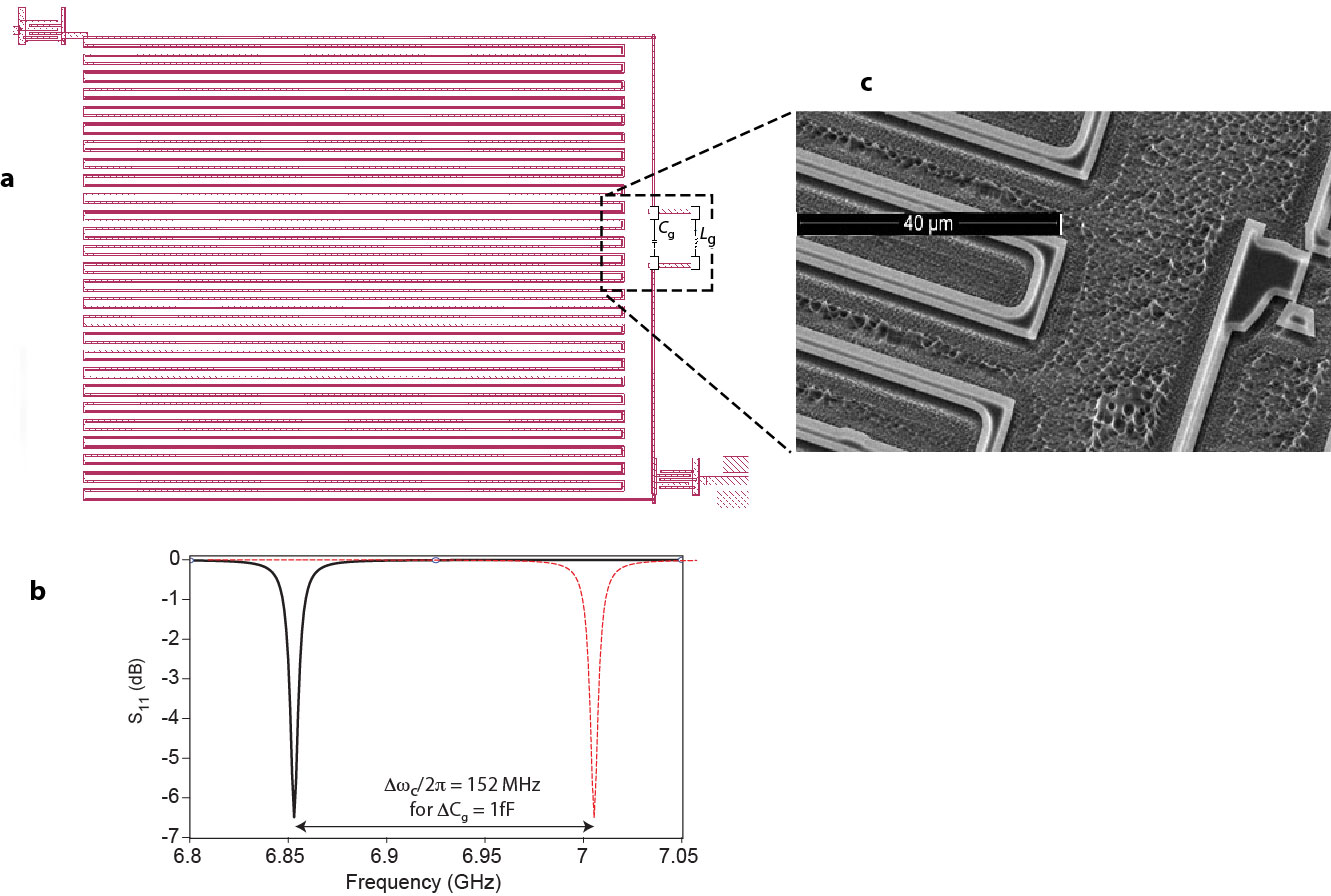}
\caption{\textbf{Design for the cavity.} \textbf{a}, Simulation drawing
  for the meandering cavity structure, with ideal circuits components
  inserted between the open ends; \textbf{b}, change of the cavity
  resonance for 1 fF change of $C_g$; \textbf{c}, micrograph showing the
  clamped beam and part of the cavity. The roughly isotropic etch
  causes about 700 nm undercut also for the
  cavity.} \label{cavitydesign}
\end{figure}

The cavity design (Fig.~\ref{cavitydesign}A) is a 2 microns wide, 45
mm long meandering microstrip floating from both ends, and the mode we
are using is the lowest mode of the structure which roughly
corresponds to $\lambda/2$ resonance in a typical transmission line
resonator (where the cross-coupling between adjacent meanders is
negligible). There are similar interdigital coupling capacitors $C_c
\simeq 6$ fF in both ends, however, only one of them is used, while
the other one is shorted to ground. In order to deduce the value of
$C$, and the validity of the parallel $LC$ model in the somewhat
complicated structure whatsoever, we made electromagnetic simulations
with ideal inductor and capacitor components $C_g$ and $L_g$ inserted
between the open ends of the meander, see
Fig.~\ref{cavitydesign}a,b. By inspecting how their values affect the
mode frequency, one can extract $C$ and $L$ from $\omega_c = [\LL(L \|
L_g \RR) \LL(C + C_g \RR)]^{-1/2}$ (ignoring the effects of $C_c$, 
$C_{S1}$ and $C_{S2}$). The effective stray capacitance, which sets the coupling
energy, is then roughly the parallel sum $C + C_c \simeq 24$ fF.

The values of $C_S$ are further determined from a lumped element
circuit simulation, by comparing to the measured reflection
parameters. From the experiment, we obtain the FWHM of the S$_{11}$
magnitude of $(2 \pi) \times 6.0$ MHz, and the maximum absorption at
resonance of -5.5 dB. Note that there is no accurate simple relation
for obtaining $\gamma_c = \omega_c / Q_c$ from the \emph{reflection}
measures. Instead, it holds for the \emph{driven response}
$\chi(\omega)$ of the $LC$ circuit that FWHM $\Delta \omega_c =
\sqrt{3} \gamma_c$. We determine $\Delta \omega_c$ from the lumped
element simulation: internal losses, modeled by a resistor, are first
adjusted to match the measured absorption. The driven response
function, including $\Delta \omega_c$, is then given by, for example,
the current $i_L$ flowing through $L$, as a function of frequency:
\begin{equation} \label{eq:chi} \chi(\omega) \equiv i_L(\omega)
  /i_L(\omega_c) = \frac{\gamma_c \omega_c}{ \sqrt{\gamma_c^2\omega^2
      + \LL(\omega^2 - \omega_c^2\RR)^2}}
\end{equation}
This way, we obtain $\Delta \omega_c \simeq (2 \pi) \times 12.2$ MHz,
and finally an estimate $\gamma_{c, S11} = (2 \pi) \times 7.0$
MHz. The ratio of internal and external dissipation $\gamma_{I}
/\gamma_{E} \simeq 3.4$ is determined by the resonance absorption.

In section \ref{sec:detune}, using pump detuning measurement, we make
the most accurate measurement to yield the final numbers $\gamma_{c} =
(2 \pi) \times 6.2$ MHz, $\gamma_I = (2 \pi) \times 1.4$ MHz and
$\gamma_E = (2 \pi) \times 4.8$ MHz, which come close to those deduced
here.

The cavity number of quanta $n_c$ at a given detuning is given by
$n_c(\omega)/n_c(\omega_c) = \chi^2(\omega)$, and $n_c(\omega_c) = L
i_L(\omega_c)^2$. The current response $i_L(\omega_c)$ at a given
input power is again obtained from lumped element simulation.

\subsection{Cryogenic setup}
\label{sec:setup}

\begin{figure}[!h]
\centering
\includegraphics[width=0.5\textwidth]{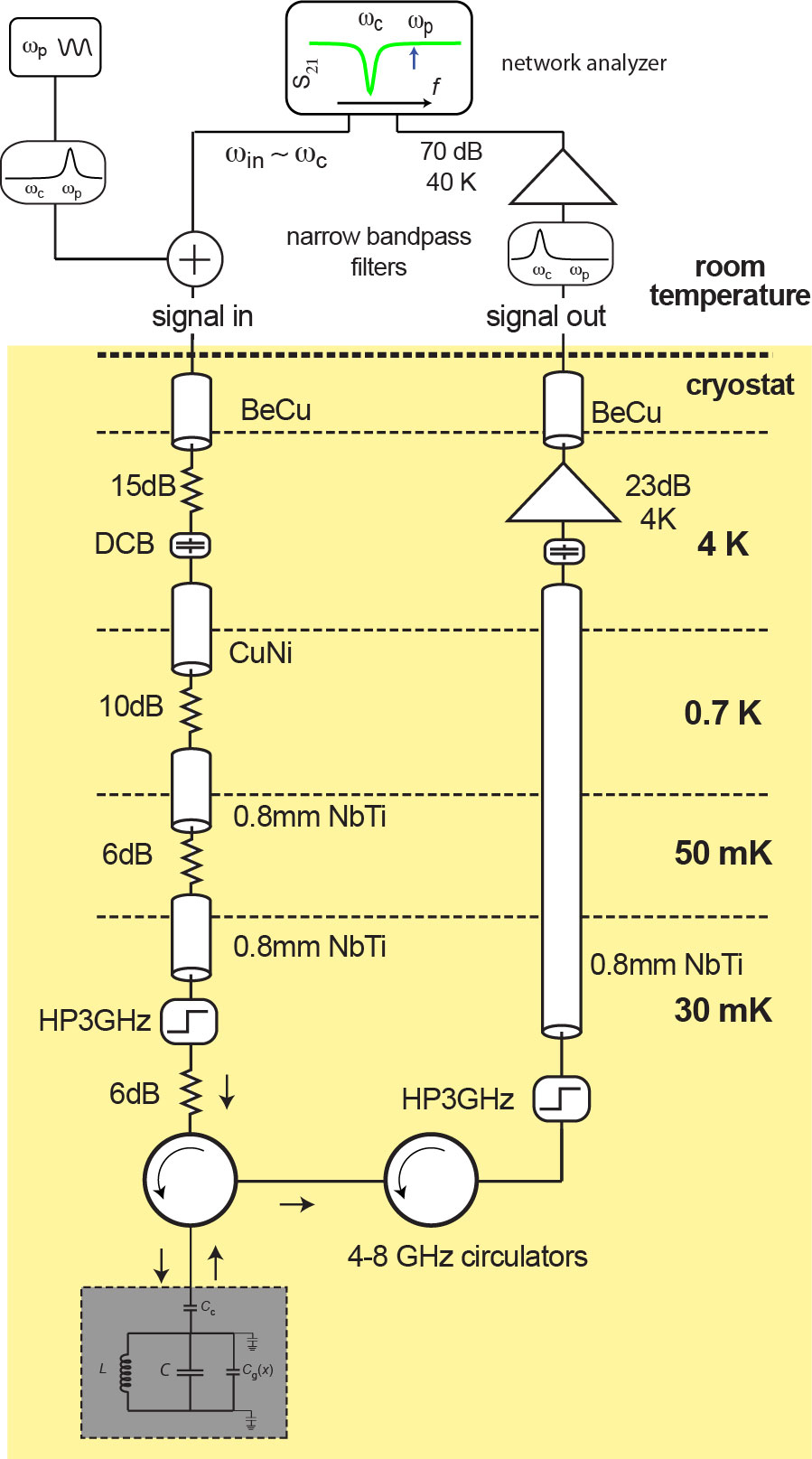}
\caption{\textbf{Setup for electronics and the microwave cabling}
  inside the dilution refrigerator for the electromechanical amplifier
  experiment. Inside the dilution cryostat, we use beryllium copper
  (BeCu), copper nickel (CuNi), and niobium titanium (NbTi) coaxial
  cables. Inner-pin DC-blocks (DCB), and high-pass filters (HP) are
  used to reduce heat leak. Back at room temperature, the pump is
  blocked from the output signal. After further amplification, the
  signal microwave is recorded coherently in a network
  analyzer.} \label{cryosetup}
\end{figure}

The experiments were carried out in a dilution refrigerator down to 25
mK temperatures. The pump and probe signals are combined at room
temperature using a power splitter. Before the signals enter the
cryostat, a sharp high-pass filter at room temperature is used to cut
the phase noise of the generators near the cavity frequency. This
filter provides 50 dB more attenuation at the cavity frequency than at
the blue sideband. Without proper filtering, the phase noise would
reflect from the cavity, and appear as extra added noise of tens of
quanta. Inside cryostat, the incoming signals are attenuated by $43
\pm 1.5$ dB. The uncertainty in the cryogenic attenuation sets the
relatively large error bars for $n_c$. Thermal noise emanating from
higher temperatures is estimated to contribute less than 0.1 quanta of
thermal occupancy into the cavity, and is thus a negligible
contribution to the total noise. The entire setup is described in
Fig.~\ref{cryosetup}.

The signals reflected from the amplifier chip are directed to the
cryogenic amplifier which has a high input compression point of -20
dBm which allows for using high pump powers without problems of
amplifier saturation. The amplifier has a noise temperature $\sim 4$
Kelvins. In addition, there is attenuation of 2...2.5 dB due to
circulators and cables between the sample and the amplifier. The
effective noise temperature, which sets the signal-to-noise ratio, is
then 6...7 K.

\subsection{Characterization of the electromechanical system}
\label{sec:detune}

In order to establish a good understanding of the basic behavior of
the electromechanical system, we determined its parameters
independently of the amplification measurements.

For determining the electromechanical coupling energy $g =
\frac{w_c}{2 C} \frac{\partial C_g}{\partial x}$, we used the value
for $C \simeq 24$ fF as obtained in section
\ref{sec:cavity}. Moreover, $\frac{\partial C_g}{\partial x} \simeq
13$ nF/m is estimated from the dimensions of the beam and the vacuum
slit. We get $g = (2 \pi) \times 1.8$ MHz/nm, which corresponds to
shift of the cavity frequency of 40 Hz per phonon.  Similarly as
previously done in Refs.~\cite{Teufel:2008te,Rocheleau:2010jd}, we made
measurements where the pump frequency or power is varied near the blue
sideband. This alters the optical spring effect which can be compared
to the theory for shifts for frequency and damping,
Eqs.~\ref{eq:geff},~\ref{eq:weff} in the supporting online text. The
effective mechanical frequency may be read from the position of
mechanical sideband, more precisely, from the departure of this peak
from the pump frequency.

\begin{figure}[!ht]
\centering
\includegraphics[width=0.4\textwidth]{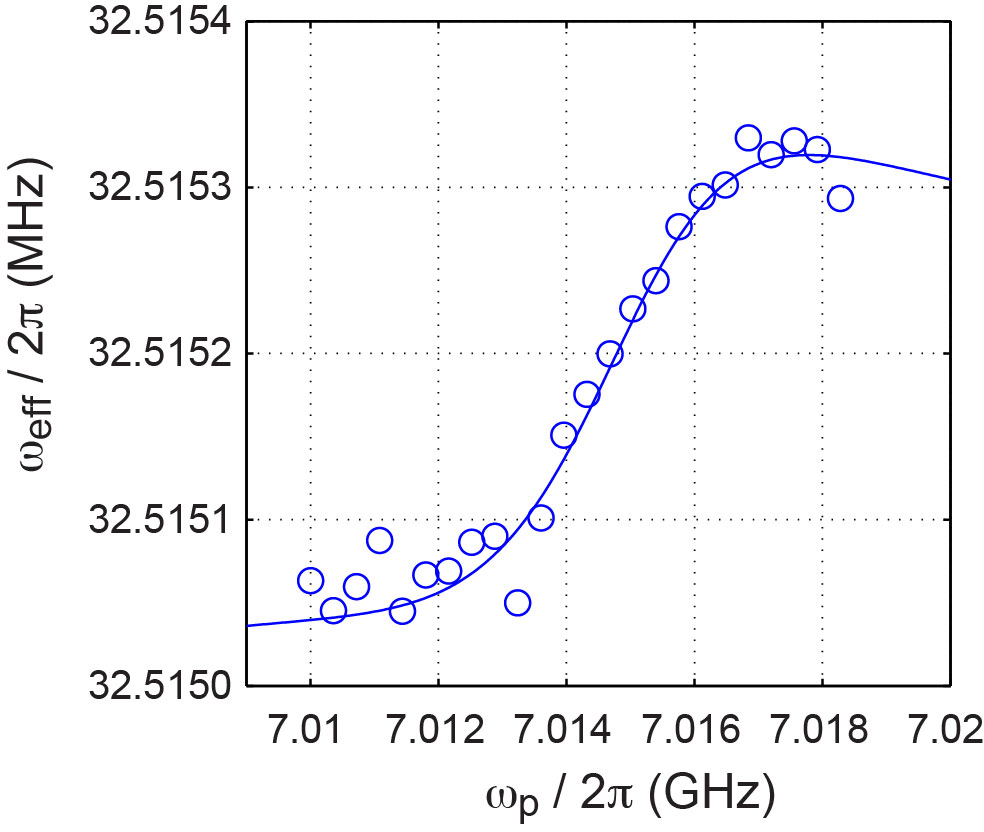}
\caption{\textbf{Characterization of the electromechanical system via
    the optical spring effect. }The incident microwave power was kept
  constant such that at the blue sideband frequency, $n_c \sim 5.3
  \times 10^5$.} \label{detunes}
\end{figure}

In Fig.~\ref{detunes} we compare the measured effective mechanical
frequency to the theory. Corresponding plots for the damping are shown
in the main text in Figs.~2A and 3D. The best fit is obtained with
$\gamma_c = 6.2$ MHz. This value differs 10 \% from that deduced from
the S$_{11}$ measurement. We consider this value of $\gamma_c$ the
most reliable, and will use it in the rest of the paper.

The values of $n_c$ we get from these fits are about 30 \% smaller
than those from independent estimates based on the input attenuation
and cavity response. We attribute this difference to the somewhat
inaccurately known cryogenic attenuation, which has a sensitive effect
on $n_c$. We adjust the scale of $n_c$ according to these fits, and
quote the adjusted values in the paper. For instance, a useful fixed
point is the instability point, which is expected according to theory
at $n_c \sim 1.2 \times 10^6$ in the situation of Fig.~3B in the
paper.

\subsection{Determination of the noise added by the mechanical amplifier}

The noise temperature of an amplifier is determined by comparing its
noise to a known noise source. Here, the noise floor which establishes
the signal-to-noise ratio, is set by the effective noise temperature
of the system, approximately 6...7 K.

\begin{figure}[!ht]
 \centering
 \includegraphics[width=0.9\textwidth]{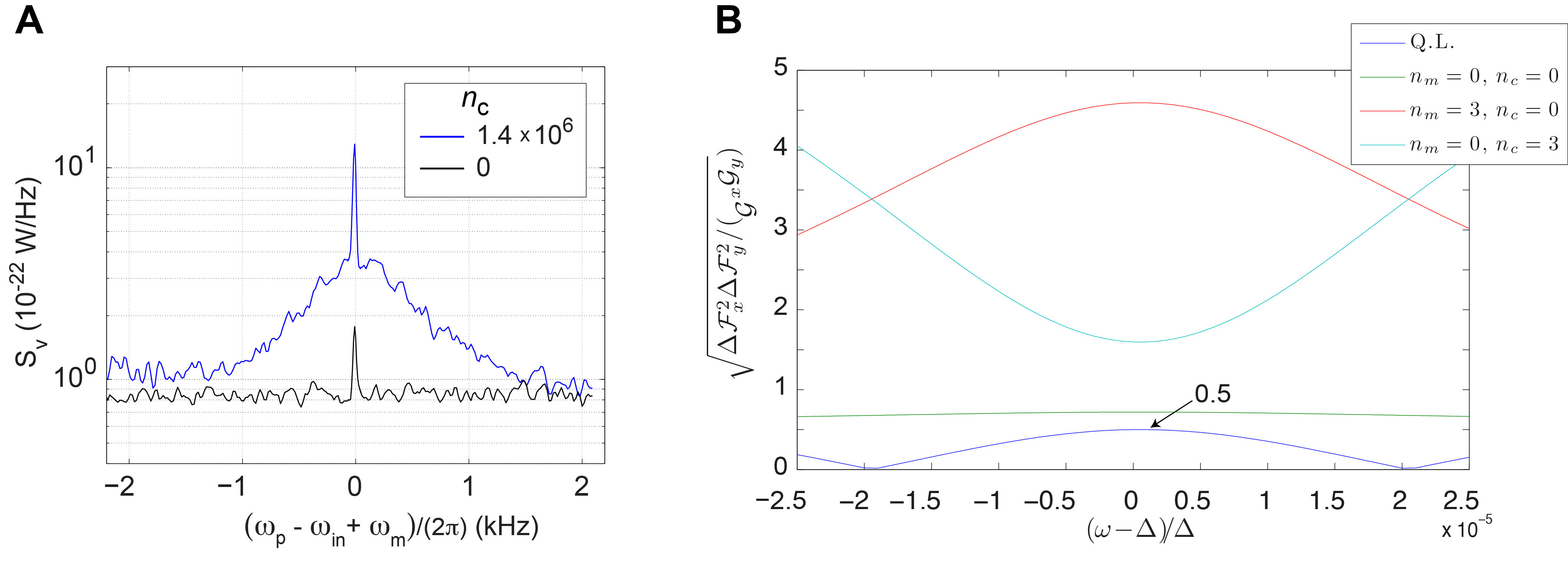}
 \caption{\textbf{Added noise of the mechanical amplifier} \textbf{a},
   A weak probe signal (narrow peak) is employed in order to deduce
   signal-to-noise ratio with the amplification off (black), or on
   (blue). $\Delta \simeq 0.89$, temperature $T = 30$ mK. The bump
   about the probe peak is due to the thermomechanical vibrations;
   \textbf{b}, theoretical plot of the added noise at the optimal value
   of the effective coupling for different values of $n_m$ and $n_c$.}
 \label{fig:noise}
\end{figure}

We worked at a temperature of 30 mK, and used a weak input signal as a
marker, see Fig.~\ref{fig:noise}. The marker peak height versus noise
floor is improved by 2.3 dB by the mechanical amplification, however,
this has to be subtracted by the cavity absorption (here, -1.7 dB). We
thus obtain a slight 0.6 dB improvement to the signal-to-noise ratio,
which corresponds to 20 added noise quanta, matching the expectation
equaling the thermal phonon number.

\section{Theoretical details}
\label{sec:theo}

\subsection{Quantum Langevin equation for the optomechanical system}
\label{sec:QLE}

In this section we derive the dynamical equations for the cavity and
the mechanical degrees of freedom for our system.  After defining the
Hamiltonians describing the two oscillators and the parametric
coupling, we write the (non-linear) Hamilton's equations for the
system. Following a standard dynamical-system approach, we separate
the dynamical variables into stationary values (in the proper rotating
frame) and corresponding fluctuations. In particular, the solutions of
the dynamical equations for the stationary values allow to determine
the value of the cavity field as a function of the pump field. These
solutions set the ``operating point'' of the amplifier, fixing the
values of the effective parameters for the fluctuations dynamics.

We now explicitly derive the quantum Langevin equations (QLE) for the
optomechanical system. In absence of any coupling to the external
world, the system (oscillator+cavity) Hamiltonian can be written as
\begin{equation}
  \label{eq:H_sys}
  H_{sys}=\hbar \omega_c(a_T^\dagger a_T + \frac{1}{2}) + H_{ho} + H_{int},
\end{equation}
where $a_T$ is the cavity field operator, and $\omega_c$
the cavity resonant frequency,
 $$
  H_{ho}=\frac{p_T^2}{2m}+ \frac{1}{2}  m \omega_m x^2
 $$
 is the mechanical harmonic oscillator Hamiltonian, $m$ the mass of
 the mechanical system and $\omega_m$ its resonant frequency. The
 Hamiltonian
 \begin{equation}
   \label{eq:H_int}
   H_{int}=-\hbar g \left( a_T^\dagger a_T+\frac{1}{2} \right)x
 \end{equation}
 is the parametric interaction Hamiltonian, where $g$ is the coupling
 between the mechanical degrees of freedom and the cavity. The
 Hamiltonian coupling the cavity with the external radiation modes can
 be written as
 \begin{equation}
   \label{eq:H_{rc}}
   H^{(I,E)}_{rc}= i \hbar \int_{-\infty}^{\infty} d \omega s_{(I,E)}(\omega)
          \left[{b_{(I,E)}}^\dagger(\omega)a_T-{b_{(I,E)}}(\omega)a_T^\dagger \right],
 \end{equation}
 where $s_{\left(I,E\right)}$ describes the cavity/reservoirs
 coupling, and the indexes $I$, $E$ refer to the external and internal
 baths respectively.  The external bath is  associated with the
 transmission line coupling the input and output signal with the cavity,
 while the internal ones refer to any other source of dissipation
 potentially coupling to the cavity.

 The reservoir associated with the dissipative dynamics of the
 mechanical oscillator (hereafter mechanical bath) can be written as
 \begin{equation}
   \label{eq:H_{mech}}
   H_{mech}= \frac{1}{2}\sum_j\left[\left(p_j-k_j x \right)^2 +\omega_j^2 q_j^2\right]
 \end{equation}
 $H_{mech}$ corresponds thus to describing the reservoir in terms of a
 collection of independent harmonic oscillators with frequencies
 $\omega_j$, with each of which is coupled to the mechanical
 oscillator through $k_j$ \cite{Caldeira:1981bf}.

 With the aid of the input-output formalism \cite{Walls:1105914}, the
 evolution equations for the cavity field operators, the position and
 momentum operators for the mechanical system can be written as
\begin{align}
    \label{eq:L_x}
    &\dot{x}=\frac{p_T}{m} \\
     \label{eq:L_p}
    &\dot{p}_T=-m \omega_m^2 x + \hbar g a_T^\dagger a_T - \gamma_m
    p_T + \xi_T\\
     \label{eq:Lang_a}
    &\dot{a}_T=-i\left(\omega_c-\omega_p-g x
    \right)a_T-\frac{\gamma_c}{2}a_T-\sqrt{\gamma_I}a^I_{in}-\sqrt{\gamma_E}a_{T\,in}.
  \end{align}
  We have here considered a situation where the cavity is strongly
  driven by a coherent field oscillating at frequency $\omega_p$.
  Moreover $\gamma_E$, $\gamma_I$ represent the losses associated with
  the input/output port and the photon bath associated with the internal
  losses of the cavity ($\gamma_c=\gamma_I+\gamma_E$), and $\gamma_m$
  the mechanical losses. We now linearize
  Eqs. (\ref{eq:Lang_a}-\ref{eq:L_p}), rewriting $a_T$, $x$ and
  $p_T$ as the sum of a coherent field and a quantum operator
\begin{align}
   &a_T=\alpha+ a \label{eq:at}\\
   &x=\chi+ \sqrt{\frac{\hbar}{m \omega_m}}q \\
   &p_T=\pi+ \sqrt{\hbar m \omega_m}p.
 \end{align}
 More specifically, we have rewritten Eq. (\ref{eq:at}) with a view to
 the decomposition in terms of a (coherent) pump field $\alpha_{\rm
   P}$ and input signal and noise sources, i.e. $$
 {a_T}_{in}=\alpha_p+a_{in}$$

 Since we are interested in the steady-state solution, neglecting all
 fluctuations, we impose the condition
 $\dot{\alpha}=\dot{\chi}=\dot{\pi}=0$, leading to the steady-state
 values (in a frame rotating at $\omega_p$)
\begin{align}
   \label{eq:st_p}
   &\pi_s=0 \\
   \label{eq:st_x}
   &\chi_s= \frac{\hbar g}{m \omega_m^2} \left(\left| \alpha_s \right|^2 + \frac{1}{2} \right) \\
   \label{eq:st_a}
&\alpha_s=\frac{\sqrt{\gamma_E} \alpha_{p}}{\frac{\gamma_c}{2}+i\left(\omega_c-\omega_p - g \chi_s \right)}.
 \end{align}
 Eqs. (\ref{eq:st_p},~\ref{eq:st_a}), can be combined into a
 third-order algebraic equation, leading to three stationary solutions
 or the cavity field $\alpha_s$ as a function of the pump field
 $\alpha_{p}$.

We  now focus on the solution for which $\alpha_{s}\to 0$ when
$\alpha_{in}\to 0$. In this case, the evolution equations for the
fluctuation operators can be written as
\begin{align}
  \label{eq:dq}
     \dot{q}&= \omega_m  p \\
  \label{eq:dp}
     \dot{p}&= - \omega_m  q - \gamma_m p + G \delta X + \xi \\
  \label{eq:da}
  \dot{a}&=i \Delta a - \frac{\gamma_c}{2} a+ \frac{G}{\sqrt{2}} q +
  \sum_{i=I,E}\sqrt{\gamma_{i}} a^{i}_{in}
\end{align}
where $\Delta=\omega_p-\omega_c-g\chi_s$, $G=2 g\sqrt{\frac{\hbar}{2 m
    \omega_m}} \alpha_s$, having assumed, without loss of generality,
that $\alpha_s$ is real. Equations (\ref{eq:dq}-\ref{eq:da}) represent the
quantum Langevin equations for the cavity+mechanical resonator system.
It is worth noting here that, following \cite{Giovannetti:2001vx}, we
have not performed the rotating wave approximation for the mechanical
bath degrees of freedom, this choice will affect the expression for
the noise spectrum for the operator $\xi$.

\subsection{Amplification}
\label{sec:ampli}

\begin{figure}[!ht]
 \centering
 \includegraphics[width=0.6\textwidth]{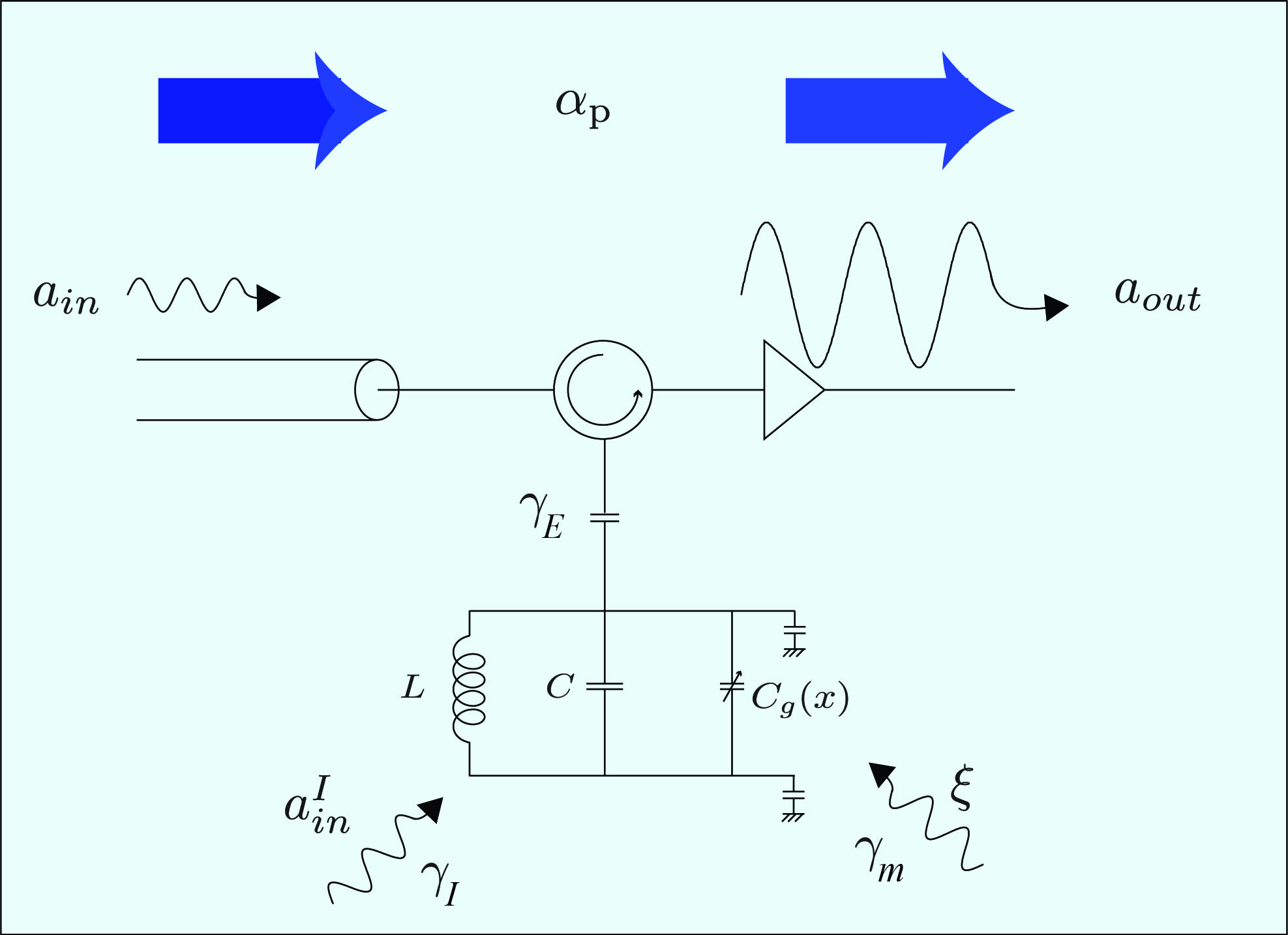}
 \caption{Schematics of the amplification scheme with an outline of
   the different noise sources. $a_{in}=\alpha_{in}+\delta a_{in}$:
   input field associated with the input signal $\alpha_{in}$ and the
   noise at the input port $\delta a_{in}$. $a_{in}^I$: field
   associated with the noise reservoir acting directly on the resonant
   cavity. $\xi$: mechanical noise associated with the thermal
   bath.} \label{fig:noisyschema}
\end{figure}

Considering the relation between input and output fields at the
input/output port of the cavity \cite{Walls:1105914}
\begin{align}
  \label{eq:io}
  a_{out}=\sqrt{\gamma_E}a+a_{in}
\end{align}
the solution of eqs. (\ref{eq:dq}~\ref{eq:da}), leads to the general
relating the output field to the various incoming fields
\begin{align}
  \label{eq:ph_sens_ampli}
  a_{out}(\omega)=M(\omega)a_{in}(\omega)+La_{in}^\dagger(\omega)+
              M_I(\omega)a^I_{in}(\omega)+L_I{a^I}_{in}^\dagger(\omega)+
              Q(\omega) \xi(\omega)
\end{align}
where $a^I_{in}(\omega)$ and $\xi(\omega)$ represent the noise
introduced by the internal losses of the cavity and the mechanical
bath (see Fig.\ref{fig:noisyschema}). The power gains for the input signal
($M$ and $L$), and those for the input noise ($M_I$, $L_I$ and $Q$) are given as 
\begin{align}
  &M(\omega)=\left[\Gamma_M(\omega)\frac{\gamma_E}{\gamma_c/2-i\left(\omega+\Delta\right)}
                    -i\frac{\gamma_E\left(\Gamma_M(\omega)-1\right)}{2\Delta}-1\right]
                \label{eq:MLQ1}  \\
  &L(\omega)=-i\frac{\gamma_E(\Gamma_M(\omega)-1)}{2 \Delta} \label{eq:MLQ2} \\
  &M_I(\omega)=\left[\Gamma_M(\omega)\frac{\sqrt{\gamma_E \gamma_I}}{\gamma_c/2-i\left(\omega+\Delta\right)}
                    -i\frac{\sqrt{\gamma_E
                        \gamma_I}\left(\Gamma_M(\omega)-1\right)}{2\Delta}\right]
                   \label{eq:MLQ3} \\
  &L_I(\omega)=-i\frac{\sqrt{\gamma_E \gamma_I}(\Gamma_M(\omega)-1)}{2 \Delta} \label{eq:MLQ4} \\
  &Q(\omega)=\sqrt{\frac{\gamma_c}{2}}\frac{\Gamma_M(\omega)-1}{\Delta G}
                  \left[\left(\Delta-\omega\right)-i\frac{\gamma_c}{2}\right]\label{eq:MLQ5}.
\end{align}
The key role in the amplification is played by the factor
\begin{align}
  \label{eq:GM}
  \Gamma_M(\omega)&=\frac{\omega_m^2-\omega^2-i \gamma_m \omega}{\omega_{\rm
      eff}^2-\omega^2-i \gamma_{\rm eff}\omega}
\end{align}
which, in turn, depends on the effective resonant frequency
\begin{align}
  \label{eq:geff}
   \omega_{\rm eff}=\left[\omega_m^2+ \frac{G^2 \Delta \omega_m \left[\gamma_c^2/4-\omega^2+\Delta^2\right]}
                        {\left[\gamma_c^2/4+\left(\omega-\Delta\right)^2\right]
                         \left[\gamma_c^2/4+\left(\omega+\Delta\right)^2\right]}\right]^{1/2}
\end{align}
and the effective damping coefficient
\begin{align}
  \label{eq:weff}
  \gamma_{\rm eff}=\left[\gamma_m - \frac{2 \gamma_c G^2 \Delta \omega_m}
                                 {\left[\gamma_c^2/4+\left(\omega-\Delta\right)^2\right]
                                  \left[\gamma_c^2/4+\left(\omega+\Delta\right)^2\right]}
          \right]
\end{align}
induced by the coupling with the cavity on the mechanical
resonator. At resonance $\omega \simeq \omega_m$ and neglecting the
(weak) pump dependence of $\omega_{\rm eff}$, it is clear that a decrease of
$\gamma_{\rm eff} \to 0$ will lead to $\Gamma_M \gg 1$ (see below for stability
considerations), and thus, through the $\Gamma_M$ dependence of $M$ and
$L$, to an amplification of an input signal.

We use the notion of preferred quadratures \cite{Caves:1982wq}, for
which the output quadrature fields $X_{out}$, $Y_{out}$ are
independent of $Y_{in}$ and $X_{in}$ respectively. The power gains in these
quadratures are obtained as
\begin{align}
  \label{eq:gainX}
  &\mathcal{G}_x=\left(\left| M \right|+\left| L \right| \right)^2 \\
   \label{eq:gainY}
  &\mathcal{G}_y=\left(\left| M \right|-\left| L \right| \right)^2 \\
   \label{eq:gainav}
  &\mathcal{G}_{av}=\frac{1}{2}\left(\mathcal{G}_x+\mathcal{G}_y\right)=\left| M \right|^2+\left| L \right|^2
\end{align}
The expression of the gain in the preferred quadratures corresponds to
the possibility of choosing an appropriate phase for $a_{in}$ and
$a_{out}$ leading to real-valued expressions for $M$ and $L$, given by
Eqs. (\ref{eq:MLQ1},~\ref{eq:MLQ2}). In these quadratures, the
amplifier equations can be written as (dropping here the added-noise
terms)
\begin{align}
  \label{eq:gai_quadr}
  &X_{out}=\left(\left| M \right|+\left| L \right|\right) X_{in} \\
  &Y_{out}=\left(\left| M \right|-\left| L \right|\right) Y_{in} \\
\end{align}
thus leading to the relations given by
Eqs. (\ref{eq:gainX},~\ref{eq:gainY}) for $\mathcal{G}_x$ and
$\mathcal{G}_y$. In addition to the trivial difference associated with
the condition $\gamma_E \neq \gamma_I$, reflecting different coupling
mechanisms of the cavity to the external world, the expressions for
$M_I(\omega)$ and $M(\omega)$ differ due to the interference term
appearing in the expression of $M(\omega)$ (the third term on the
right-hand side of Eq. \eqref{eq:MLQ1}). This term represents the
interference between the input (either signal or noise) that has been
reflected at the cavity/transmission line interface and the input that
has travelled through the cavity. The expression of the gains given by
Eqs. (\ref{eq:gainX}-\ref{eq:gainav}) involve the coefficients
$M(\omega)$ and $L(\omega)$ only, since, obviously, the signal is
supposed to enter the system through the input external port only.  On
the other hand, while opening the cavity to the transmission line will
also open that port to the noise from the transmission line, this
noise is regarded as intrinsic noise of the input signal and thus does
not contribute to the noise added by the amplifier.  The system will
thus be open to one signal source (the coherent part of $a_{in}$), the
noise associated with internal losses ($a{^I}_{in}$), the noise
associated with the mechanical bath ($\xi$) and the noise from the
transmission line (the incoherent part of $a_{in}$), the latter not
contributing to the noise added by the amplifier.

\subsection{Noise: input field correlators and the quantum limit}
\label{sec:noise}

Within this scheme, the noise added by the amplifier can be expressed
in terms of noise spectra associated with the internal losses and the
mechanical bath. Following a standard approach \cite{Genes:2008hr}, the
correlators for $a_{in}$ and $a^I_{in}$ are given by
\begin{align}
  \label{eq:corr_ain}
  &\langle a^{(I)}_{in}(t) {a^{(I)}_{in}}^\dagger(t') \rangle
  =\left[n(\omega_c)+1\right]\delta(t-t') \\
  &\langle  {a^{(I)}_{in}}^\dagger(t)a^{(I)}_{in}(t') \rangle
  =n(\omega_c)\delta(t-t').
\end{align}

Similarly, the mechanical noise correlator can be written as \cite{Clerk:2010dh}
\begin{align}
  \label{eq:corr_xi}
  \langle\xi(t)\xi(t')\rangle= \int \frac{d\omega}{2\pi} \exp\left[-i\omega(t-t')\right] S_\xi(\omega)
\end{align}
with
\begin{align}
    \label{eq:xi_nsp}
      S_\xi(\omega)=2 \gamma_m \frac{\omega}{\omega_m}
          \left\{(n_\omega+1) \Theta(\omega)+\left[(n_{-\omega}+1) \Theta(-\omega)\right]\right\}
\end{align}
where $\Theta(x)$ is the Heaviside step function (see
e.g. \cite{Clerk:2010dh}). Considering a thermally populated bath, the
noise spectrum assumes the form \cite{Genes:2008hr}
\begin{align}
  \label{eq:xicorr_th}
  S_\xi(\omega)=\gamma_m
  \frac{\omega}{\omega_m}\left[ \coth\left(\frac{\hbar \omega}{kT}\right)+1\right]
\end{align}.

We are now in the position to evaluate the noise added by the
amplifier. We here define the operators
\begin{align}
  \label{eq:addnoiseX}
   \mathcal{F}_x=\frac{1}{\sqrt{2}}\left[M_I a^I_{in} + L_I
     {a^I_{in}}^\dagger+Q \xi + h.c. \right]  \\
  \label{eq:addnoiseY}
   \mathcal{F}_y= \frac{-i}{\sqrt{2}}\left[M_I a^I_{in} + L_I
     {a^I_{in}}^\dagger+Q \xi - h.c.\right]
\end{align}
where the appropriate phase has been included in the definition of
$M_I$, $L_I$ and $Q$ in order to satisfy the condition $M,L \in
\mathbb{R}$.  $(\Delta \mathcal{F}_x)^2$ and $(\Delta
\mathcal{F}_y)^2$ represent the added noise by the amplifier
\cite{Caves:1982wq}. The condition establishing a lower bound for the
added noise reads in this case
\begin{align}
  \label{eq:qlim}
   \sqrt{\left|\Delta \mathcal{F}_x \right|^2\left|\Delta
      \mathcal{F}_y \right|^2/\left(\mathcal{G}_x\mathcal{G}_y\right)}
  \geq \frac{1}{4} \left|1-\left(\mathcal{G}_{x}\mathcal{G}_{y}\right)^{-1}\right|.
\end{align}
Close to the optimal effective coupling $G_{opt}=\sqrt{\gamma_m \gamma_c}$, and
for $\omega  \simeq \omega_m$, the expression for the added noise is
given by
\begin{align}
  \label{eq:addnoise_s}
   \sqrt{\left|\Delta \mathcal{F}_x \right|^2\left|\Delta
      \mathcal{F}_y
    \right|^2/\left(\mathcal{G}_x\mathcal{G}_y\right)}\simeq
    \frac{\gamma_I}{\gamma_c} (n^I_{opt}+1/2)+\frac{\gamma_c}{\gamma_E}(n_m+1/2).
\end{align}
From Eq. \eqref{eq:addnoise_s}, it is possible to see that the quantum
limit for the amplification can be reached in absence of internal
cavity losses and for a zero-temperature mechanical reservoir. In our
experimental setup $n^I_{opt}\simeq 0$, leading to a linear increase of
the added noise with the number of mechanical reservoir phonons. The
linear dependence coefficient is given by the ratio between total and
external losses.

\subsection{Stability and validity of the linearized model}
\label{sec:stab}

In obtaining the QLE for the cavity and the mechanical resonator, we
have linearized the equations of motion for the cavity+mechanical
resonator system. We will here discuss the criterion for the stability
and the limits of validity of the (linear) QLE equations considered to
analyze the system dynamics (Ginsburg criterion) \cite{Mancini:1994vf}.
The requirement for the system stability is that the poles of the
effective mechanical susceptibility, induced by the coupling between
the mechanical resonator and the cavity, lie in the lower complex
half-plane. In other terms, the effective mechanical damping
$\gamma_{\rm eff}$ must be positive in order for the system to be stable. The
condition $\gamma_{\rm eff} \to 0^+$ correspond to the situation of maximal
gain and, on crossing the $\gamma_{\rm eff}=0$ value, to the loss of
stability.
\begin{figure}[h!b]
\centering
 \includegraphics[width=0.7\textwidth]{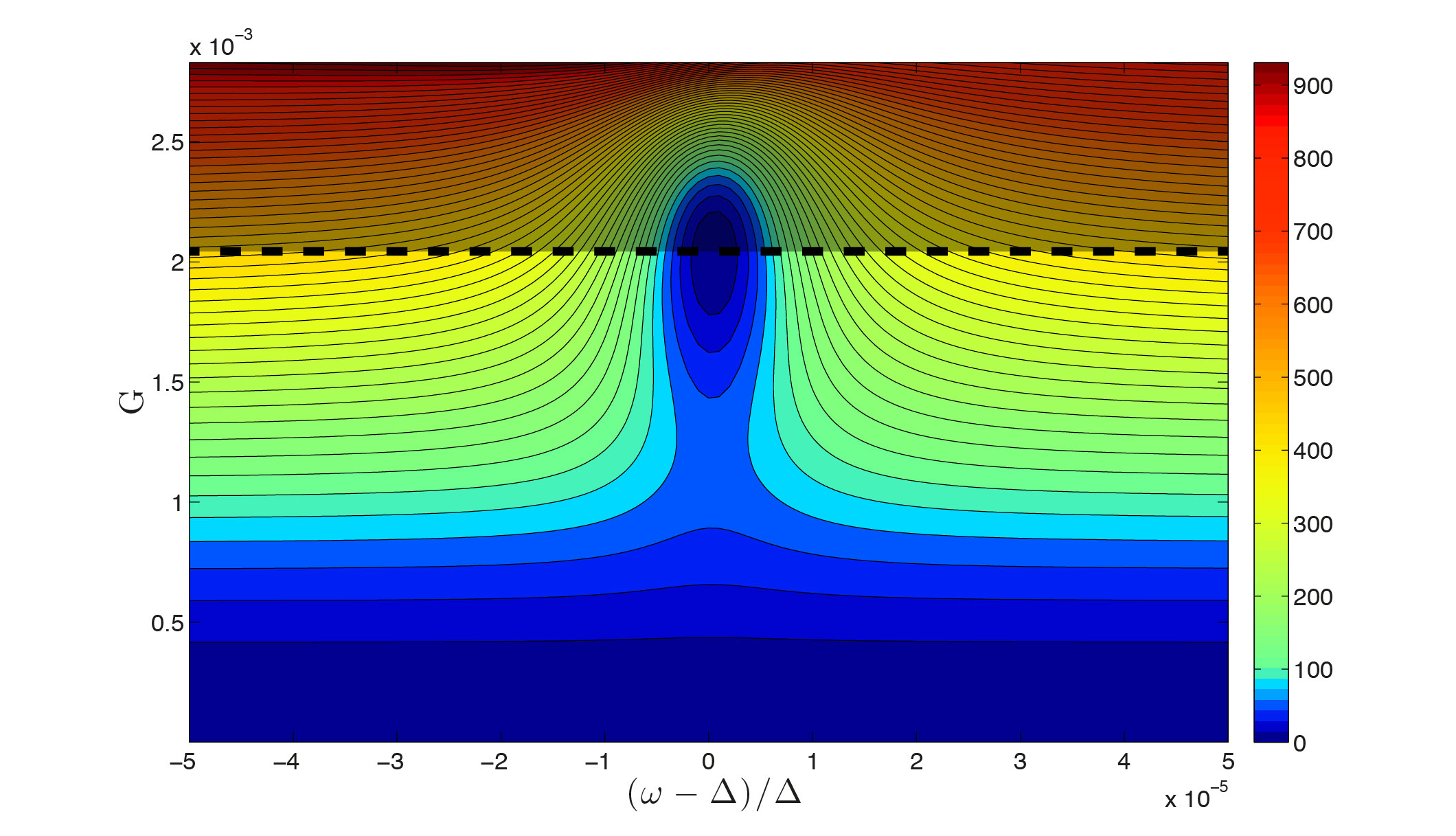}
 \caption{Number of signal photons ensuring the validity of the
   linear-regime analysis as a function of $\omega$ and $G$. We have
   assumed $\left(\langle
     a^\dagger_{in}a_{in}\rangle/\alpha_S^2\right)_{\rm threshold} =
   10^{-4}/|\Gamma_M|^2$}
  \label{fig:lin_lim}
\end{figure}
In the linearization procedure we have assumed that the term $x \cdot a_T$
appearing in Eq. ~\eqref{eq:Lang_a} could be expanded as
\begin{align}
  \label{eq:linxa}
  (\xi+q)\cdot(\alpha+a) \simeq \xi \alpha+ \alpha q + \xi a
\end{align}
analogously,
\begin{align}
  \label{eq:linada}
   a_T^\dagger a_T \simeq \alpha^2 + \alpha^* a + \alpha a^\dagger.
\end{align}
Eqs. \eqref{eq:linxa} and \eqref{eq:linada} thus establish that, for
the linearized QLE equations to aptly describe the dynamics of the
optomechanical system, the following conditons must be met
\begin{align}
  \label{eq:cond_x}
  \frac{\langle q a \rangle}{\alpha_s \chi_s} \ll 1 \\
  \label{eq:cond_a}
   \frac{\langle a^\dagger a \rangle}{\alpha_s^2} \ll 1,
\end{align}
The solutions for $q$ and $a$ of the QLE as a function of the input
field $a_{in}$ lead to the following condition for the ratio between
the signal and the pump power
\begin{align}
  \label{eq:sign_to_pump}
   \frac{\langle a^\dagger_{in}a_{in}\rangle}{\alpha_S^2} \ll \frac{1}{|\Gamma_M|^2}.
\end{align}

It is thus clear from Eq. ~\eqref{eq:sign_to_pump} that for large
enough values of $|\Gamma_M|$, the linearized description of the
system physics breaks down. However as it can be seen from
Fig.~\ref{fig:lin_lim}, there is a large range of parameters where,
while having a gain significantly larger than 1, the linear model is
still valid.

\bibliographystyle{naturemag}